# SwinCross: Cross-modal Swin Transformer for Head-and-Neck Tumor Segmentation in PET/CT Images

Gary Y. Li, Junyu Chen, Se-In Jang, Kuang Gong, and Quanzheng Li

*Abstract*—Radiotherapy (RT) combined with cetuximab is the standard treatment for patients with inoperable head and neck cancers. Segmentation of head and neck (H&N) tumors is a prerequisite for radiotherapy planning but a time-consuming process. In recent years, deep convolutional neural networks have become the de facto standard for automated image segmentation. However, due to the expensive computational cost associated with enlarging the field of view in DCNNs, their ability to model long-range dependency is still limited, and this can result in sub-optimal segmentation performance for objects with background context spanning over long distances. On the other hand, Transformer models have demonstrated excellent capabilities in capturing such long-range information in several semantic segmentation tasks performed on medical images. Inspired by the recent success of Vision Transformers and advances in multi-modal image analysis, we propose a novel segmentation model, debuted, Cross-Modal Swin Transformer (SwinCross), with cross-modal attention (CMA) module to incorporate cross-modal feature extraction at multiple resolutions. To validate the effectiveness of the proposed method, we performed experiments on the HECKTOR 2021 challenge dataset and compared it with the nnU-Net[1] (the backbone of the top-5 methods in HECKTOR 2021) and other state-of-the-art transformer-based methods such as UNETR[2], and Swin UNETR[3]. The proposed method is experimentally shown to outperform these comparing methods thanks to the CMA module's ability to capture better inter-modality complimentary feature representations between PET and CT, for the task of head-and-neck tumor segmentation.

## I. Introduction

HEAD and Neck (H&N) cancers are among the most common cancers worldwide [4], accounting for about 4% of all cancers in the United States. FDG-PET and CT imaging are the gold standards for the initial staging and follow-up of H&N cancer. Quantitative image biomarkers from medical images such as radiomics have previously shown tremendous potential to optimize patient care, particularly for Head and Neck tumors [5]. However, radiomics analyses rely on an expensive and error-prone manual process of annotating the Volume of Interest (VOI) in 3D. The automatic segmentation of H&N tumors from PET/CT images could therefore enable the validation of radiomics models on very large cohorts and with optimal reproducibility. Besides, automatic segmentation algorithms could enable a faster clinical workflow. By focusing on metabolic and anatomical features respectively, PET and CT include complementary and synergistic information in the context of H&N primary tumor segmentation.

Recently Transformer, a neural network based on self-attention mechanisms to compute feature representations and global dependencies, has flourished in natural language processing and computer vision [6]. In computer vision, Transformer-based architectures have achieved remarkable success and have demonstrated superior performance on a variety of tasks, including visual recognition [7, 8], objection detection [9, 10], semantic segmentation [11, 12], etc. [8, 13-15]. The success of vision transformers in the computer vision field has inspired their use in medical imaging, where they have shown promising potential in various applications, such as classification [16-18] segmentation [2, 19, 20] and registration [21, 22]. Chen et al. first proposed the TransUNet [19] for medical image segmentation, which used a 12-layer ViT for the bottleneck features and followed the 2D UNet design and adopted the Transformer blocks in the middle structure. Later

This work was supported by NIH under Grant NIH C06 CA059267. (*Corresponding authors: Gary Y. Li*)

Gary Y. Li is with Philips Research North America, Cambridge, MA 02141, (e-mail: bettergary@gmail.com). Work done while a postdoc at The Center for Advanced Medical Computing and Analysis, Massachusetts General Hospital/Harvard Medical School, Boston, MA 02114

Se-In Jang, Kuang Gong and Quanzheng Li are with The Center for Advanced Medical Computing and Analysis, Massachusetts General Hospital/Harvard Medical School, Boston, MA 02114 (e-mail:

sjang7@mgh.harvard.edu; kgong@mgh.harvard.edu; quanzheng@mgh.harvard.edu).

Junyu Chen is with The Russell H Morgan Department of Radiology and Radiological Science, School of Medicine, Johns Hopkins University, Baltimore, MD 21287 (e-mail: jchen245@jhmi.edu ).

that year, two improved versions of TransUNet, TransUNet+[23], and Ds-TransUNet [24], were proposed and achieved better results for CT segmentation tasks. For 3D segmentation where the computational cost for self-attention becomes very expensive, researchers have attempted to limit the use of transformer blocks, i.e., only use self-attention at the bottleneck between the encoder and decoder network [25, 26] or adopted a deformable mechanism which enables attention on a small set of key positions [27]. SegTran[28] proposed to leverage the learning tradeoff between larger context and localization accuracy by doing pairwise feature contextualization with squeeze and excitation blocks. More recently, more and more state-of-the-art performance has been refreshed by networks with pre-trained transformer backbone. Pre-training techniques have become a new area of research in transformers as the self-attention blocks commonly require pre-training data at a large scale to learn a more powerful backbone [29]. For example, self-supervised Swin UNETR (Tang et al., 2021) collects a large-scale of CT images (5,000 subjects) for pre-training the Swin Transformer encoder, which derives significant improvement and state-of-the-art performance for BTCV [30] and Medical Segmentation Decathlon (MSD) [31]. Self-supervised masked autoencoder (MAE) [32] investigates the MAE-based self-pre-training paradigm designed for Transformers, which enforces the network to predict masked targets by collecting information from the context. Besides developing advanced architectures to better learn the data, researchers have also attempted to improve performance by providing additional data that is more specific to the task the network is given.

In representation learning, the advancement of multimodal learning has benefited numerous applications [33, 34]. The utilization of fused features from multimodalities has largely improved performance in cross-media analysis tasks such as video classification [35], event detection [36, 37], and sentiment analysis [38, 39]. A characteristic that these works have demonstrated in common is that better features for one modality (e.g., audio) can be learned if multiple modalities (e.g., audio and video) are present at feature learning time. In [40], Ngiam et al proposed the cross-modality (audio + video) feature learning scheme for shared representation learning and demonstrated superior visual speech classification performance compared to the classifier trained with audio-only or video-only data. Wang et al. proposed a DNN-based model combining canonical correlated autoencoder and autoencoder-based terms to fuse multi-view for an unsupervised multi-view feature learning [41]. Following this trend, deep learning-based multimodal methods have also gained attraction in the medical image analysis community due to their remarkable performance in many medical image analysis tasks including the classification [42, 43], diagnosis [44, 45], image-retrieval [46], and segmentation [47-49]. Carneiro et al. [50] proposed to use of shared image features from unregistered views of the same region to improve classification performance. In [44], Xu et al. proposed to jointly learn the nonlinear correlations between image and other non-image modalities for cervical dysplasia diagnosis by leveraging multimodal information, which significantly outperformed methods using any single source of information alone. In [45], Suk et al. proposed to learn a joint feature representation from MRI and PET using a hierarchical DCNN for Alzheimer's Disease diagnosis.

Despite the impressive representation capacity of vision transformer models, current vision transformer-based segmentation models still suffer from inconsistent and incorrect dense predictions when fed with multi-modal input data. We suspect that the power of their self-attention mechanism is limited in extracting the complementary information exisit in multi-modal data. To this end, we propose a dual-branch cross-attention Swin Transformer (SwinCross) to combine image patches from two different modalities at different scale to produce more complementary feature representations from the two modalities. Furthermore, to reduce computation, we develop a cross-modal attention (CMA) module based on cross attention and the shifted window self-attention mechanism from Swin Transformer [10]. To validate the effectiveness of the proposed method, we performed experiments on a public dataset and compared the proposed method with state-of-the-art methods such as UNETR, Swin UNETR, and nnU-Net. The proposed method is experimentally shown to be able to capture the inter-modality correlation between PET and CT for the task of head-and-neck tumor segmentation.

## II. RELATED WORK

### A. The Current State-of-the-art Methods for H&N Tumor Segmentation

The top-five performing teams in the HECTOR 2021 challenge all used U-Net or its variants for the primary H&N tumor segmentation task [51]. In [52], Xie et al. used a patch-based 3D nnU-Net with Squeeze and Excitation normalization and a novel training scheme, where the learning rate is adjusted dynamically using polyLR [53]. The approach achieved a 5-fold average Dice score of 0.764 on the validation data dataset, which ranked them first on the leaderboard for the tumor segmentation task. They trained five models in a five-fold cross-validation manner with random data augmentation including rotation, scaling, mirroring, Gaussian noise, and Gamma correction. The final test results were generated by ensembling five test predictions via probability averaging. In [54], An et al. proposed a coarse-to-fine framework using a cascade of three U-Nets. The first U-Net is used to coarsely segment the tumor and then select a bounding box. Then, the second U-Net performs a finer segmentation on the smaller region within the bounding box, which has been shown to often lead to more accurate segmentation [55]. Finally, the last U-Net takes as input the concatenation of PET, CT, and the previous segmentation to refine the predictions. The three U-Nets were trained with different objectives – the first one to optimize the recall and the rest two to optimize the Dice score. The final results were obtained via majority voting on three different predictions: an ensemble of five nnU-Nets, an ensemble of three U-Nets with squeeze-and-excitation (SE) normalization, and the predictions from the proposed model. In [56], Lu et al. proposed a huge ensemble learning model which consists of fourteen 3D U-Nets, including the eight models adopted in [57], winner of the HECTOR 2020 challenge, five models trained with leave-on-center-out, and one model combining a prior and posteriori attention. The final ensembled prediction was generated by averaging all fourteen predictions and thresholding the resulting mask to 0.5. In [58], Yousefirizi et al. used a 3D nnU-Net with SE normalization trained on a leave-one-center-out with a combination of a "unified" focal and Mumford-Shah losses, leveraging the advantage of distribution, region, and boundary-based loss functions. Lastly, Ren et al [59] proposed a 3D nnU-Net with various PET normalization techniques, namely PET-clip and PET-sin. The former clips the Standardized Uptake Values (SUV) range in [0,5] and the latter transforms monotonic spatial SUV increase into onion rings via a sine transform of SUV, which ranked them fifth on the leaderboard. Although CNN-based methods have outstanding representation capability, it is lacking the ability to model long-range dependencies due to the limited receptive fields of the convolution kernels. This inherent limitation of receptive size sets an obstacle to learning global semantic information which is critical for dense prediction tasks like segmentation.

### B. Transformers and Multi-modal Learning

Transformers have been widely applied in the fields of Natural Language Processing [60-62] and Computer Vision [63-68] primarily due to its excellent capability to model long-range dependency. Besides achieving impressive performance in a variety of language and vision tasks, the Transformer model also provides an effective mechanism for multi-modal reasoning by taking different modality inputs as tokens for self-attention[69-78]. For example, Prakash et al.[74] proposed to use a Transformer to integrate image and LiDAR representations using attention. Going beyond language and vision, we propose to utilize a cross-modal attention Swin Transformer to fuse 3D PET and CT images at multiple resolutions for the segmentation of H&N tumors. We build the SwinCross architecture based on the shifted window block from Swin Transformer, which only computes self-attention within local regions, unlike conventional ViTs, which are more computationally expensive. Although Swin Transformer is unable to explicitly compute correspondences beyond its field of view, similar to how ConvNets operate to some extent, the shifted window mechanism still yields much larger kernels than most ConvNets [79].

## III. SWINCROSS

### A. Overall Architecture of SwinCross

In this work, we propose an architecture for 3D multi-modal segmentation with two main components: (1) a Cross-modal Swin Transformer for integrating information from multiple modalities (PET and CT), and (2) a cross-modal shifted window attention block for learning complementary information from the modalities. Our key idea is to exploit the cross-modal attention mechanism to incorporate the global context for PET and CT modalities given their complementary nature for the H&N tumor segmentation task. We illustrate the architecture of SwinCross in Fig. 1. The input image to the SwinCross model is multi-channel 3D volume $F^{in} \in R^{H \times W \times D \times M}$, with a dimension of $H \times W \times D \times M$. The input image is first split channel-wise, forming a set of single-channel 3D images $F_{mod\_1}, \dots, F_{mod\_k} \in R^{H \times W \times D}$. Then, we split each single-channel image into small non-overlapped patches with a patch size of $\frac{H}{H'} \times \frac{W}{W'} \times \frac{D}{D'}$, which corresponds to a patch resolution of $H' \times W' \times D'$. Each 3D patch is projected into an embedding space with dimension C to form a tokenized sequence $S_{mod\_k} \in R^{N \times C}$, where $N = H' \times W' \times D'$ is the number of tokens in the sequence and each token is represented by a feature vector of dimensionality $C$. The $S^{mod\_K}$ sequences are inputs to the encoder network.

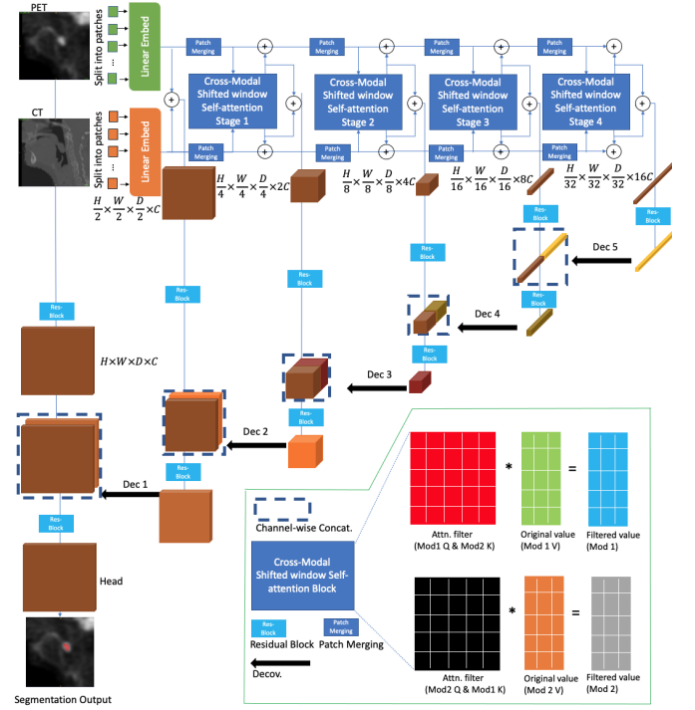

Fig. 1. Architecture of SwinCross. 3D PET and CT volumes are used as inputs to our Cross-modal attention Swin Transformer (SwinCross) which adopts multiple cross-modal attention (CMA) modules for the fusion of intermediate feature maps between the two modalities. To effectively combine patch tokens from both modalities at different scales, we develop a fusion method based on the CMA blocks, which exchange information between two branches at multiple resolutions ($\frac{1}{4}, \frac{1}{8}, \frac{1}{16}$, and $\frac{1}{32}$ of the input resolution) throughout the two feature extracting branches resulting in 5 feature vectors ($\frac{1}{2}, \frac{1}{4}, \frac{1}{8}, \frac{1}{16}$, and $\frac{1}{32}$ of the input resolution) from both modalities, which are combined via element-wise summation. The 5 feature vectors constitute fused representations of the CT and PET image at 5 different resolutions. These feature vectors are then processed with a ConvNet decoder which predicts the final segmentation map. We channel-wise concatenate the decoded feature vectors from a previous resolution to the feature vector at the current resolution and use the resulting feature vectors as input to the deconvolution block to produce the feature vector at the next resolution.

### B. Network Encoder

The encoder uses linear projections for computing a set of queries, keys and values ($Q$, $K$, and $V$) for each input sequence $S_{mod\_k}$.

$$Q_{mod\_k} = S_{mod\_k} M^q, K_{mod\_k} = S_{mod\_k} M^k, V_{mod\_k} = S_{mod\_k} M^v \qquad (1)$$

where $M^q \in R^{D_f \times D_q}$, $M^k \in R^{D_f \times D_k}$, and $M^v \in R^{D_f \times D_v}$ are weight matrices. In the case of bimodal cross-attention, it uses the scaled dot products between the $Q$ and $K$ of each modality to compute the attention weights and then aggregates the values for each query of each modality,

$$A_{mod\_1} = softmax\left(\frac{Q_{mod_1}K_{mod_2}^T}{\sqrt{D_k}}\right)V_{mod_1}, \qquad (2)$$

$$A_{mod\_2} = softmax\left(\frac{Q_{mod\_2}K_{mod\_1}^T}{\sqrt{D_k}}\right)V_{mod\_2}, \qquad (3)$$

in which $Q_{mod\_1}$, $K_{mod\_1}$, $V_{mod\_1}$, $Q_{mod\_2}$, $K_{mod\_2}$, $V_{mod\_2}$ denote queries, keys, and values from modality 1 and 2, respectively; $D_k$ represents the size of the key and query.

As these are 3D tokens and the attention computation cost increases quadratically with respect to the number of tokens, we adopted the shifted window mechanism for the cross-attention calculation. Specifically, we utilize windows of size $M \times M \times M$ to evenly partition the patchified volume into $\frac{H'}{M} \times \frac{W'}{M} \times \frac{D'}{M}$ regions at a given layer $l$ in the transformer encoder. In the subsequent layers of $l$ and $l + 1$ of the encoder, the outputs are calculated as

$$\hat{A}_{mod\_k}^l = \text{W-MSA}\left(\text{LN}\left(A_{mod_k}^{l-1}\right)\right) + A_{mod_k}^{l-1} \qquad (4)$$

$$A_{mod_k}^l = \text{MLP}\left(\text{LN}\left(\hat{A}_{mod\_k}^l\right)\right) + \hat{A}_{mod\_k}^l \qquad (5)$$

$$\hat{A}_{mod\_k}^{l+1} = \text{SW-MSA}\left(\text{LN}\left(A_{mod_k}^l\right)\right) + A_{mod_k}^l \qquad (6)$$

$$A_{mod_k}^{l+1} = \text{MLP}\left(\text{LN}\left(\hat{A}_{mod\_k}^{l+1}\right)\right) + \hat{A}_{mod\_k}^{l+1} \qquad (7)$$

A 3D version of the cyclic-shifting [10] was implemented for efficient computation of the shifted window mechanism. SwinCross follows a standard four-stage structure [10] but has cross-modality attention mechanism at each stage for fusion of intermediate feature maps between both modalities. The fusion is applied at multiple resolutions ($\frac{H}{2} \times \frac{W}{2} \times \frac{D}{2} \times C, \frac{H}{4} \times \frac{W}{4} \times \frac{D}{4} \times 2C, \frac{H}{8} \times \frac{W}{8} \times \frac{D}{8} \times 4C, \frac{H}{16} \times \frac{W}{16} \times \frac{D}{16} \times 8C$) throughout the feature extractor of both modalities resulting in four filtered feature maps ($\frac{H}{4} \times \frac{W}{4} \times \frac{D}{4} \times 2C, \frac{H}{8} \times \frac{W}{8} \times \frac{D}{8} \times 4C, \frac{H}{16} \times \frac{W}{16} \times \frac{D}{16} \times 8C, \frac{H}{32} \times \frac{W}{32} \times \frac{D}{32} \times 16C$) from each modality. The filtered feature maps from both modalities are summed element-wise and sent to the decoder, as indicated by the plus signs on Fig. 1. At each stage, these feature maps are fed back into each of the individual modality branches using an element-wise summation with the down-sampled (via patch merging) input feature maps, as indicated by the green plus signs on Fig. 1.

The SwinCross encoder has a patch size of $2 \times 2 \times 2$ and a feature dimension of $2 \times 2 \times 2 \times 2 = 16$, taking into account the multi-modal PET/CT images with 2 channels. The size of the embedding space C is set to 48 in our encoder. Furthermore, the SwinCross encoder has 4 stages which comprise of $[2,4,2,2]$ cross-modal shifted window transformer blocks at each stage. Hence, the total number of layers in the encoder is L = 10. Before stage 1, each single-channel image was split into small non-overlapped patches by a 3D convolution layer with stride size equal to 2 (patch size) and output channels equal to C, resulting in $\frac{H}{2} \times \frac{W}{2} \times \frac{D}{2} \times C$ 3D tokens. To follow the hierarchical structure proposed in [10], a patch merging layer is used on each modality branch to decrease the resolution of the feature representations by a factor of 2 at the beginning of each stage. In order to preserve fine details from the input image to the output segmentation, we send the original input multi-channel 3D volume $F^{in} \in R^{H \times W \times D \times M}$ and its embedded version together with the feature map outputs from the 4 stages to the decoder, resulting in a total of 6 feature maps with dimensions of $H \times W \times D \times M$, $\frac{H}{2} \times \frac{W}{2} \times \frac{D}{2} \times C$, $\frac{H}{4} \times \frac{W}{4} \times \frac{D}{4} \times 2C$, $\frac{H}{8} \times \frac{W}{8} \times \frac{D}{8} \times 4C$, $\frac{H}{16} \times \frac{W}{16} \times \frac{D}{16} \times 8C$, and $\frac{H}{32} \times \frac{W}{32} \times \frac{D}{32} \times 16C$.

## C. Network Decoder

We adopted a ConvNet decoder as opposite to a Transformer decoder for the ease of cross-modal feature fusion and lower computational cost. SwinCross adopts a U-shaped network design in which the extracted feature representations of the encoder are used in the decoder via skip connections at each resolution. At each stage $i$ ($i \in [0,1,2,3,4,5]$) of the encoder, the output feature representations are reshaped into size $\frac{H}{2^i} \times \frac{W}{2^i} \times \frac{D}{2^i}$ and fed into a residual block comprising of two 3x3x3 convolutional layers that are normalized by instance normalization layers. Subsequently, the resolution of the feature maps is increased by a factor of 2 using a deconvolutional layer and the outputs are concatenated with the outputs of the previous stage. The concatenated features are then fed into another residual block as previously described. The final segmentation outputs are computed by using a 1x1x1 convolutional layer and a sigmoid activation function.

## IV. RESULTS AND DISCUSSION

### A. Ablation Studies on HECTOR 2021 Dataset

In Table 1, we ablate the CMA module block, which only concerns the attention mechanism of the Swin Transformer, and we keep everything else the same as in Swin UNETR (e.g., embed dimension, feature size, number of blocks in each stage, window size, and number of heads). We start from channel-wise concatenated input, which consists of two volume images from both modalities. This multi-modal input already gives Swin UNETR a strong 5-fold average Dice Score of $0.754\pm0.032$. If we send in the images from two modalities in two separate branches (as shown in Fig. 1) and use CMA module block to fuse the learned features from each modality at each stage, the performance is improved to $0.769\pm0.026$. The output from each CMA module block has the same shape as the input and each filtered feature is added back to the corresponding modality's branch. At each stage, the sum of the filtered features from the CMA module block is sent to the decoder.

Table 1. Ablation study of CMA module on HECKTOR 2021 dataset

| Model | Block composition | Embed Dimension | Feature Size | Number of Blocks | Window Size | Number of Heads | 5-fold Average Dice Score |
|---|---|---|---|---|---|---|---|
| Swin UNETR | W-MSA+ SW-MSA | 768 | 48 | [2,2,2,2] | [7,7,7] | [3,6,12,24] | 0.754 ±0.032 |
| | CMW-MSA + CMSW-MSA | 768 | 48 | [2,2,2,2] | [7,7,7] | [3,6,12,24] | **0.769 ±0.026** |

## B. Comparison to the State-of-the-art Methods in Medical Image Segmentation

We have compared the performance of SwinCross against the current SOTA methods in medical image segmentation such as Swin UNETR, UNETR, and nnU-Net, using a 5-fold cross-validation split. Evaluation results (dual modality) across all five folds are presented in Table 2. The proposed SwinCross model achieved the highest 5-fold average Dice score of 0.769 among all the comparing methods. Note that SwinCross outperformed Swin UNETR across all 5 folds, which demonstrated its capability of learning multi-modal feature representations at multiple resolutions via the Cross-modal attention modules. These results are consistent with our previous findings [80], in which we showed that nnU-Net outperformed Swin UNETR for H&N tumor segmentation on two public datasets. With the CMA block and dual-branch fusion mechanism, SwinCross demonstrates a slightly better segmentation performance than nnU-Net, as measured by the 5-fold average Dice score. However, competitive performance is seen from nnU-Net, which again indicates that for small object segmentation, the improvement from modeling long-range dependency may be limited as a smaller effective field may be enough to capture all the foreground and background information of the small object such as a H&N tumor [80].

Table 2. Five-fold cross-validation benchmarks in terms of Dice score values from all methods using PET and CT image.

| Dice Score | SwinCross (PET+CT) proposed | Swin UNETR (PET+CT) | nnU-Net (PET+CT) | UNETR (PET+CT) |
|---|---|---|---|---|
| Fold0 | **0.717** | 0.715 | 0.714 | 0.702 |
| Fold1 | **0.788** | 0.781 | 0.781 | 0.716 |
| Fold2 | 0.800 | 0.752 | **0.803** | 0.727 |
| Fold3 | **0.779** | 0.772 | 0.777 | 0.762 |
| Fold4 | **0.761** | 0.748 | **0.761** | 0.708 |
| Average | **0.769** | 0.754 | 0.767 | 0.723 |

Fig. 2 illustrates sample segmentation outputs of all the methods. For large tumors, Fig. 2a shows the benefit of modeling long-range dependency brought by the transformer-based models. SwinCross was the only network that could capture the tip of the tumor marked by the yellow crosshair, demonstrating the benefit of the CMA module blocks, which allow feature exchange between two modalities at multiple resolutions in the encoder. For smaller tumors, Fig. 2c shows that SwinCross was able to capture the fine edge of the tumor by incorporating complementary edge features from CT image, outperforming the other methods that used channel-wise concatenated input. The results confirmed the findings in [34] – performing fusion within the network is better than outside the network.

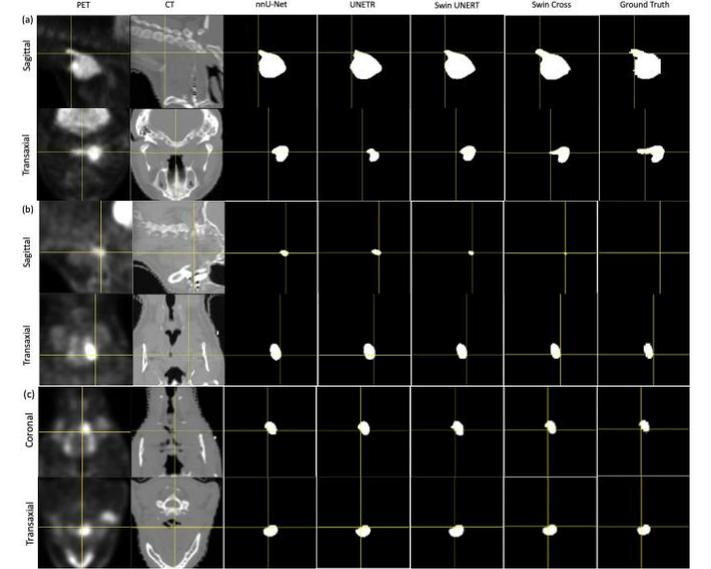

Fig.2. From left to right are input PET image, CT image, inferenced mask from nnU-Net, UNETR, Swin UNETR, SwinCross (proposed) and ground truth.

## C. Comparison to Single-modality Segmentation

As a comparison to using dual-modality input, we have computed the performance of the reference methods using single-modality input with the same 5-fold cross-validation split. Since H&N tumor is primarily present in PET, we conducted these single-modality experiments using PET image only. Evaluation results (PET only) across all five folds are presented in Table 3. The Swin UNETR achieved the highest 5-fold average Dice score of 0.732 among the reference methods for single-modality input. Fig. 3 shows box plot of the mean dice score values of the five splits from all the methods using singe-modality as well as dual-modality. Overall, the plot demonstrates that networks with dual-modality (PET and CT) input significantly outperforms the same networks with single-modality (PET) input for the task of H&N tumor segmentation.

Table 3. Five-fold cross-validation benchmarks in terms of mean Dice score values from the reference methods using PET image only.

| Dice Score | UNETR (PET) | Swin UNETR (PET) | nnU-Net (PET) |
|---|---|---|---|
| Fold0 | 0.668 | 0.689 | 0.683 |
| Fold1 | 0.677 | 0.711 | 0.701 |
| Fold2 | 0.715 | 0.768 | 0.762 |
| Fold3 | 0.722 | 0.762 | 0.745 |
| Fold4 | 0.686 | 0.729 | 0.728 |
| Average | 0.694 | 0.732 | 0.724 |

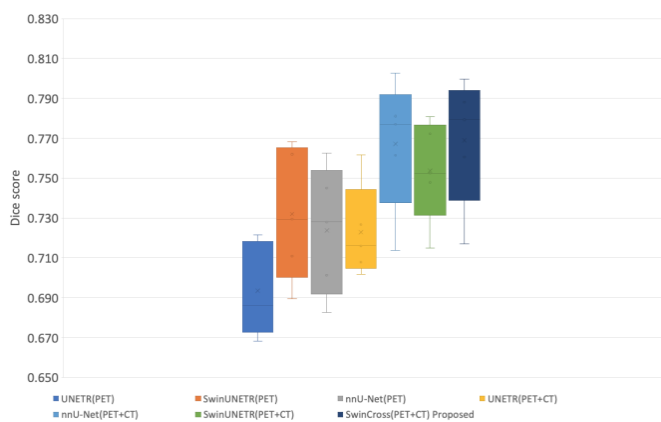

Fig. 3. Box plot for the mean dice score values of the five splits for different methods.

## V. CONCLUSION

A Cross-modal Swin Transformer was introduced for the automatic delineation of head and neck tumors in PET and CT images. The proposed model has a cross-modality attention module that uses feature exchange between two modalities at multiple resolutions. A ConvNet-based decoder is connected to the encoder via skip connections at different resolutions. We have validated the effectiveness of our proposed model by comparing with the state-of-the-art methods using the HECTOR 2021 dataset. The proposed method is experimentally shown to outperform the other methods by capturing better inter-modality correlation between PET and CT for the task of head-and-neck tumor segmentation. The method proposed is generally applicable to other semantic segmentation tasks using other imaging modalities such as SPECT/CT, or MRI.

## REFERENCES

1. Fabian Isensee, P.F.J., Simon A. A. Kohl, Jens Petersen and Klaus H. Maier-Hei, *nnU-Net: a self-configuring method for deep learning-based biomedical image segmentation.* Nature Methods, 2020.

2. Ali Hatamizadeh, Y.T., Vishwesh Nath, Dong Yang, Andriy Myronenko, Bennett Landman, Holger R. Roth, Daguang Xu. *UNETR: Transformers for 3D Medical Image Segmentation.* in *IEEE/CVF Winter Conference on Applications of Computer Vision (WACV).* 2022.

3. Ali Hatamizadeh, V.N., Yucheng Tang, Dong Yang, Holger R. Roth, and Daguang Xu, *Swin UNETR: Swin Transformers for Semantic Segmentation of Brain Tumors in MRI Images.* arxiv.org, 2022.

4. Parkin, D.M., et al., *Global cancer statistics, 2002.* CA Cancer J Clin, 2005. **55**(2): p. 74-108.

5. Andrearczyk, V., et al. *Overview of the HECKTOR Challenge at MICCAI 2020: Automatic Head and Neck Tumor Segmentation in PET/CT.* 2021. Cham: Springer International Publishing.

6. Li, J., et al., *Transforming medical imaging with Transformers? A comparative review of key properties, current progresses, and future perspectives.* arXiv preprint arXiv:2206.01136, 2022.

7. L. Yuan, Q.H., Z. Jiang, J. Feng and S. Yan, *VOLO: Vision Outlooker for Visual Recognition.* IEEE Transactions on Pattern Analysis and Machine Intelligence, 2022.

8. Yang, C.a.W., Yilin and Zhang, Jianming and Zhang, He and Wei, Zijun and Lin, Zhe and Yuille, Alan. *Lite Vision Transformer with Enhanced Self-Attention.* in *IEEE/CVF Conference on Computer Vision and Pattern Recognition.* 2022.

9. Wang, W.a.X., Enze and Li, Xiang and Fan, Deng-Ping and Song, Kaitao and Liang, Ding and Lu, Tong and Luo, Ping and Shao, Ling, *Pvtv2: Improved baselines with pyramid vision transformer.* Computational Visual Media, 2022.

10. Liu, Z., et al. *Swin transformer: Hierarchical vision transformer using shifted windows.* in *Proceedings of the IEEE/CVF International Conference on Computer Vision.* 2021.

11. Bowen Cheng, A.S., Alexander Kirillov. *Per-Pixel Classification is Not All You Need for Semantic Segmentation.* in *Advances in Neural Information Processing Systems.* 2021.

12. nze Xie, W.W., Zhiding Yu , Anima Anandkumar , Jose M. Alvarez , Ping Luo. *SegFormer: Simple and Efficient Design for Semantic Segmentation with Transformers.* in *Advances in Neural Information Processing Systems.* 2021.

13. Luo, Z.L.a.W.W.a.E.X.a.Z.Y.a.A.A.a.J.M.A.a.T.L.a.P. *Panoptic SegFormer: Delving Deeper into Panoptic Segmentation with Transformers.* in *IEEE/CVF Conference on Computer Vision and Pattern Recognition.* 2022.

14. Wan, Z.a.Z., Jingbo and Chen, Dongdong and Liao, Jing. *High-Fidelity Pluralistic Image Completion with Transformers.* in *IEEE/CVF Conference on Computer Vision and Pattern Recognition.* 2022.

15. Wang, H.a.Z., Yukun and Adam, Hartwig and Yuille, Alan and Chen, Liang-Chieh. *MaX-DeepLab: End-to-End Panoptic Segmentation with Mask Transformers.* in *IEEE/CVF Conference on Computer Vision and Pattern Recognition.* 2022.

16. Hou, B., et al. *Ratchet: Medical transformer for chest x-ray diagnosis and reporting.* in *Medical Image Computing and Computer Assisted Intervention–MICCAI 2021: 24th International Conference, Strasbourg, France, September 27–October 1, 2021, Proceedings, Part VII 24.* 2021. Springer.

17. Matsoukas, C., et al., *Is it time to replace cnns with transformers for medical images?* arXiv preprint arXiv:2108.09038, 2021.

18. Park, S., et al., *Federated split task-agnostic vision transformer for COVID-19 CXR diagnosis.* Advances in Neural Information Processing Systems, 2021. **34**: p. 24617-24630.

19. Chen, J.a.L., Yongyi and Yu, Qihang and Luo, Xiangde and Adeli, Ehsan and Wang, Yan and Lu, Le and Yuille, Alan L. and Zhou, Yuyin, *TransUNet: Transformers Make Strong Encoders for Medical Image Segmentation.* arXiv.org, 2021.

20. Hatamizadeh, A., et al. *Swin unetr: Swin transformers for semantic segmentation of brain tumors in mri images.* in *Brainlesion: Glioma, Multiple Sclerosis, Stroke and Traumatic Brain Injuries: 7th International Workshop, BrainLes 2021, Held in Conjunction with MICCAI 2021, Virtual Event, September 27, 2021, Revised Selected Papers, Part I.* 2022. Springer.

21. Chen, J., et al., *Transmorph: Transformer for unsupervised medical image registration.* Medical image analysis, 2022. **82**: p. 102615.

22. Chen, J., et al., *Vit-v-net: Vision transformer for unsupervised volumetric medical image registration.* arXiv preprint arXiv:2104.06468, 2021.

23. Yuhang Liu, H.W., Zugang Chen, Kehan Huangliang, Haixian Zhang, *TransUNet+: Redesigning the skip connection to enhance features in medical image segmentation.* Knowledge-Based Systems, 2022. **256**.

24. Lin, A.a.C., Bingzhi and Xu, Jiayu and Zhang, Zheng and Lu, Guangming, *DS-TransUNet:Dual Swin Transformer U-Net for Medical Image Segmentation.* arXiv.org, 2021.

25. Chang, Y., et al., *TransClaw U-Net: Claw U-Net with Transformers for Medical Image Segmentation.* ArXiv, 2021. **abs/2107.05188**.

26. Deng, K.a.M., Yanda and Gao, Dongxu and Bridge, Joshua and Shen, Yaochun and Lip, Gregory and Zhao, Yitian and Zheng, Yalin, *TransBridge: A Lightweight Transformer for Left Ventricle Segmentation in Echocardiography*, in *Simplifying Medical Ultrasound: Second International Workshop, ASMUS 2021, Held in Conjunction with MICCAI 2021, Strasbourg, France, September 27, 2021, Proceedings.* 2021. p. 63–72.

27. Xie, Y.a.Z., Jianpeng and Shen, Chunhua and Xia, Yong. *CoTr: Efficiently Bridging CNN and Transformer for 3D Medical Image Segmentation.* in *International conference on medical image computing and computer-assisted intervention.* 2021.

28. Li, S., et al., *Medical Image Segmentation using Squeeze-and-Expansion Transformers.* ArXiv, 2021. **abs/2105.09511**.

29. Dosovitskiy, A., et al., *An image is worth 16x16 words: Transformers for image recognition at scale. arXiv 2020.* arXiv preprint arXiv:2010.11929, 2010.

30. Landman, B., et al. *Miccai multi-atlas labeling beyond the cranial vault–workshop and challenge.* in *Proc. MICCAI Multi-Atlas Labeling Beyond Cranial Vault—Workshop Challenge.* 2015.

31. Antonelli, M., et al., *The Medical Segmentation Decathlon.* Nature Communications, 2022. **13**(1): p. 4128.

32. Zhou, L., et al., *Self pre-training with masked autoencoders for medical image analysis.* arXiv preprint arXiv:2203.05573, 2022.

33. Guo, W.Z., J.W. Wang, and S.P. Wang, *Deep Multimodal Representation Learning: A Survey.* Ieee Access, 2019. **7**: p. 63373-63394.

34. Guo, Z., et al., *Deep Learning-based Image Segmentation on Multimodal Medical Imaging.* IEEE Trans Radiat Plasma Med Sci, 2019. **3**(2): p. 162-169.

35. Liu, Y., X. Feng, and Z. Zhou, *Multimodal video classification with stacked contractive autoencoders.* Signal Processing, 2016. **120**: p. 761-766.

36. Wu, S., et al. *Zero-shot event detection using multimodal fusion of weakly supervised concepts.* in *Proceedings of the IEEE Conference on Computer Vision and Pattern Recognition.* 2014.

37. Habibian, A., T. Mensink, and C.G. Snoek, *Video2vec embeddings recognize events when examples are scarce.* IEEE transactions on pattern analysis and machine intelligence, 2016. **39**(10): p. 2089-2103.

38. Poria, S., et al., *Fusing audio, visual and textual clues for sentiment analysis from multimodal content.* Neurocomputing, 2016. **174**: p. 50-59.

39. Zadeh, A., et al., *Tensor fusion network for multimodal sentiment analysis.* arXiv preprint arXiv:1707.07250, 2017.

40. Ngiam, J., et al. *Multimodal deep learning.* in *ICML.* 2011.

41. Wang, W., et al., *On Deep Multi-View Representation Learning: Objectives and Optimization.* ArXiv, 2016. **abs/1602.01024**.

42. Zhu, Z.Y., et al. *Multi-View Perceptron: a Deep Model for Learning Face Identity and View Representations.* in *28th Conference on Neural Information Processing Systems (NIPS).* 2014. Montreal, CANADA.

43. Carneiro, G., J. Nascimento, and A.P. Bradley. *Unregistered Multiview Mammogram Analysis with Pre-trained Deep Learning Models.* in *18th International Conference on Medical Image Computing and Computer-Assisted Intervention (MICCAI).* 2015. Munich, GERMANY.

44. Xu, T., et al. *Multimodal Deep Learning for Cervical Dysplasia Diagnosis.* in *MICCAI.* 2016.

45. Suk, H.I., et al., *Hierarchical feature representation and multimodal fusion with deep learning for AD/MCI diagnosis.* Neuroimage, 2014. **101**: p. 569-582.

46. Kang, Y., S. Kim, and S. Choi. *Deep Learning to Hash with Multiple Representations.* in *12th IEEE International Conference on Data Mining (ICDM).* 2012. Brussels, BELGIUM.

47. Zhu, X., et al., *Medical lesion segmentation by combining multimodal images with modality weighted UNet.* Medical Physics, 2022. **49**(6): p. 3692-3704.

48. Guo, Z., et al. *MEDICAL IMAGE SEGMENTATION BASED ON MULTI-MODAL CONVOLUTIONAL NEURAL NETWORK: STUDY ON IMAGE FUSION SCHEMES.* in *15th IEEE International Symposium on Biomedical Imaging (ISBI).* 2018. Washington, DC.

49. Guo, Z., et al., *Gross tumor volume segmentation for head and neck cancer radiotherapy using deep dense multi-modality network*. Phys Med Biol, 2019. **64**(20): p. 205015.

50. Carneiro, G., J. Nascimento, and A.P. Bradley. *Unregistered multiview mammogram analysis with pre-trained deep learning models*. in *International Conference on Medical Image Computing and Computer-Assisted Intervention*. 2015. Springer.

51. Andrearczyk, V., et al. *Overview of the HECKTOR Challenge at MICCAI 2021: Automatic Head and Neck Tumor Segmentation and Outcome Prediction in PET/CT Images*. 2022. Cham: Springer International Publishing.

52. Xie, J. and Y. Peng. *The Head and Neck Tumor Segmentation Based on 3D U-Net*. 2022. Cham: Springer International Publishing.

53. Chen, L.-C., et al., *Deeplab: Semantic image segmentation with deep convolutional nets, atrous convolution, and fully connected crfs*. IEEE transactions on pattern analysis and machine intelligence, 2017. **40**(4): p. 834-848.

54. An, C., H. Chen, and L. Wang. *A Coarse-to-Fine Framework for Head and Neck Tumor Segmentation in CT and PET Images*. 2022. Cham: Springer International Publishing.

55. Zhou, Y., et al. *A Fixed-Point Model for Pancreas Segmentation in Abdominal CT Scans*. 2017. Cham: Springer International Publishing.

56. Lu, J., et al. *Priori and Posteriori Attention for Generalizing Head and Neck Tumors Segmentation*. 2022. Cham: Springer International Publishing.

57. Iantsen, A., D. Visvikis, and M. Hatt. *Squeeze-and-Excitation Normalization for Automated Delineation of Head and Neck Primary Tumors in Combined PET and CT Images*. 2021. Cham: Springer International Publishing.

58. Yousefirizi, F., et al. *Segmentation and Risk Score Prediction of Head and Neck Cancers in PET/CT Volumes with 3D U-Net and Cox Proportional Hazard Neural Networks*. 2022. Cham: Springer International Publishing.

59. Ren, J., et al. *PET Normalizations to Improve Deep Learning Auto-Segmentation of Head and Neck Tumors in 3D PET/CT*. 2022. Cham: Springer International Publishing.

60. Vaswani, A., et al. *Attention Is All You Need*. in *31st Annual Conference on Neural Information Processing Systems (NIPS)*. 2017. Long Beach, CA.

61. Devlin, J., et al., *Bert: Pre-training of deep bidirectional transformers for language understanding*. arXiv preprint arXiv:1810.04805, 2018.

62. Brown, T., et al., *Language models are few-shot learners*. Advances in neural information processing systems, 2020. **33**: p. 1877-1901.

63. Wang, X., et al. *Non-local neural networks*. in *Proceedings of the IEEE conference on computer vision and pattern recognition*. 2018.

64. Parmar, N., et al. *Image transformer*. in *International conference on machine learning*. 2018. PMLR.

65. Dosovitskiy, A., et al., *An image is worth 16x16 words: Transformers for image recognition at scale*. arXiv preprint arXiv:2010.11929, 2020.

66. Carion, N., et al. *End-to-end object detection with transformers*. in *European conference on computer vision*. 2020. Springer.

67. Chen, M., et al., *Generative Pretraining From Pixels*, in *Proceedings of the 37th International Conference on Machine Learning*, D. Hal, III and S. Aarti, Editors. 2020, PMLR: Proceedings of Machine Learning Research. p. 1691--1703.

68. Chen, C.-F.R., Q. Fan, and R. Panda. *Crossvit: Cross-attention multi-scale vision transformer for image classification*. in *Proceedings of the IEEE/CVF international conference on computer vision*. 2021.

69. Tan, H. and M. Bansal, *Lxmert: Learning cross-modality encoder representations from transformers*. arXiv preprint arXiv:1908.07490, 2019.

70. Li, L.H., et al., *Visualbert: A simple and performant baseline for vision and language*. arXiv preprint arXiv:1908.03557, 2019.

71. Sun, C., et al. *Videobert: A joint model for video and language representation learning*. in *Proceedings of the IEEE/CVF International Conference on Computer Vision*. 2019.

72. Chen, Y.-C., et al. *Uniter: Universal image-text representation learning*. in *European conference on computer vision*. 2020. Springer.

73. Li, X., et al. *Oscar: Object-semantics aligned pre-training for vision-language tasks*. in *European Conference on Computer Vision*. 2020. Springer.

74. Prakash, A., K. Chitta, and A. Geiger. *Multi-modal fusion transformer for end-to-end autonomous driving*. in *Proceedings of the IEEE/CVF Conference on Computer Vision and Pattern Recognition*. 2021.

75. Huang, Z., et al. *Seeing out of the box: End-to-end pre-training for vision-language representation learning*. in *Proceedings of the IEEE/CVF Conference on Computer Vision and Pattern Recognition*. 2021.

76. Hu, R. and A. Singh. *Unit: Multimodal multitask learning with a unified transformer*. in *Proceedings of the IEEE/CVF International Conference on Computer Vision*. 2021.

77. Akbari, H., et al., *Vatt: Transformers for multimodal self-supervised learning from raw video, audio and text*. Advances in Neural Information Processing Systems, 2021. **34**: p. 24206-24221.

78. Hendricks, L.A., et al., *Decoupling the role of data, attention, and losses in multimodal transformers*. Transactions of the Association for Computational Linguistics, 2021. **9**: p. 570-585.

79. Ding, X., et al. *Scaling up your kernels to 31x31: Revisiting large kernel design in cnns*. in *Proceedings of the IEEE/CVF Conference on Computer Vision and Pattern Recognition*. 2022.

80. Ye Li, J.C., Se-in Jang, Kuang Gong, Quanzheng Li, *Investigation of Network Architecture for Multimodal*

*Head-and-Neck Tumor Segmentation.* Arxiv.org, 2022.